\documentclass[showkeys,aps,twocolumn,showpacs,preprintnumbers,amsmath,amssymb,prd,letterpaper,floatfix,nofootinbib,superscriptaddress,]{revtex4-1}

\usepackage{amsmath,amssymb}
\usepackage[usenames,dvipsnames]{color}
\usepackage{graphicx}
\usepackage{subfigure}
\usepackage{soul}

\newcommand\bef{\begin{figure}}
\newcommand\eef[1]{\label{fg:#1}\end{figure}}
\newcommand\besf{\begin{subfigure}}
\newcommand\eesf[1]{\label{sfg:#1}\end{subfigure}}
\newcommand\beq{\begin{equation}}
\newcommand\eeq[1]{\label{#1}\end{equation}}
\newcommand\beqa{\begin{eqnarray}}
\newcommand\eeqa[1]{\label{#1}\end{eqnarray}}
\newcommand\bet{\begin{table}}
\newcommand\eet[1]{\label{tb:#1}\end{table}}
\newcommand\best{\begin{subtable}}
\newcommand\eest[1]{\label{stb:#1}\end{subtable}}
\newcommand\betb{\begin{center}\begin{tabular}}
\newcommand\eetb{\end{tabular}\end{center}}
\newcommand\beit{\begin{itemize}}
\newcommand\eeit{\end{itemize}}

\newcommand\fgn[1]{Figure \ref{fg:#1}}

\newcommand\tbn[1]{Table \ref{tb:#1}}

\begin{document}
\title{Quantum Numbers of Recently Discovered $\Omega^{0}_{c}$ Baryons from Lattice QCD}

\author{M.\ \surname{Padmanath}}
\email{Padmanath.Madanagopalan@physik.uni-regensburg.de}
\affiliation{Instit\"ut  fur  Theoretische  Physik,  Universit\"at  Regensburg,
Universit\"atsstrase  31,  93053  Regensburg,  Germany.}

\author{Nilmani\ \surname{Mathur}}
\email{nilmani@theory.tifr.res.in}
\affiliation{Department of Theoretical Physics, Tata Institute of Fundamental
         Research,\\ Homi Bhabha Road, Mumbai 400005, India.}


\begin{abstract}
We present the ground and excited state spectra of $\Omega^{0}_{c}$ baryons  
with spin up to 7/2  from lattice quantum chromodynamics with
dynamical quark fields. Based on our lattice results, we predict the quantum 
numbers of five $\Omega^{0}_{c}$ baryons, which have recently been observed 
by the LHCb Collaboration. Our results strongly indicate that the observed states 
$\Omega_c(3000)^0$ and $\Omega_c(3050)^0$ have spin-parity $J^P = 1/2^{-}$, 
the states $\Omega_c(3066)^0$ and $\Omega_c(3090)^0$ have $J^P = 3/2^{-}$, 
whereas $\Omega_c(3119)^0$ is possibly a $5/2^{-}$ state.
\end{abstract}
\maketitle

The study of heavy hadrons is passing through an incredible era with
the discovery of numerous heavy subatomic particles~\cite{PDG}. 
As a result, there has been significant
resurgence in scientific interest to explore the spectrum of strongly
interacting heavy hadrons. To add to 
this proliferated interest in hadron spectroscopy, the LHCb
Collaboration has recently reported its unambiguous observation of
five new resonances in $\Xi^{+}_{c}K^{-}$ invariant mass distribution
based on $pp$ collision data in the energy range between 
$3000$ and $3120$ MeV~\cite{Aaij:2017nav}. These resonances have been 
interpreted as the excited states of $\Omega^{0}_{c}$ baryon.
 While the masses and widths of these resonances 
 are known precisely, their other important quantum
numbers ($J^{P}$), namely, spin ($J$) and parity ($P$), are yet unknown.  
In this Letter, we predict the quantum numbers of these five $\Omega^{0}_{c}$
resonances using lattice quantum chromodynamics (lattice QCD).

On the theoretical side, potential models have been very successful in
describing regular heavy mesons. 
Using these models, several results were also reported
on heavy
baryons~\cite{Ebert:2007nw,Ebert:2011kk,Garcilazo:2007eh,Valcarce:2008dr,Roberts:2007ni,Vijande:2012mk,Yoshida:2015tia,Shah:2016nxi}.
The spectra of heavy baryons were also studied by other models, such as
QCD sum rules~\cite{Bagan:1992tp,Huang:2000tn,Wang:2009cr,Chen:2015kpa} 
and heavy
quark effective theory~\cite{Chiladze:1997ev}.  In this aspect, this
recent discovery at LHCb provides a good opportunity for testing
predictions of these models.

On the other hand, lattice QCD methods provide a unique opportunity to
study hadronic physics from first principles, particularly the energy
spectra of hadrons. Substantial progress has been made to extract the
ground and excited states of charm
mesons~\cite{Dudek:2007wv,Liu:2012ze,Moir:2013ub}.  However, most
lattice studies on heavy baryons are confined mainly to the ground
states of spin-$1/2^+$ and spin-$3/2^+$
baryons~\cite{Mathur:2002ce,Lewis:2001iz,Durr:2012dw,
  Basak:2012py,Basak:2013oya,Namekawa:2013vu,Brown:2014ena,Bali:2015lka}.
Following the successful programs in calculating the excited state
spectra of light hadrons by the Hadron Spectrum
Collaboration (HSC), recently we reported our findings on the
excited state spectra of triply charmed baryons~\cite{Padmanath:2013zfa},
doubly charmed baryons~\cite{Padmanath:2015jea}, and preliminary
results on singly charmed
baryons~\cite{Padmanath:2013bla,Padmanath:2014bxa,Padmanath:2015bra}.  Here we report
for the first time our findings on the energy spectra of $\Omega_c^0$
baryons with spin up to 7/2 for both positive and negative parity in
detail.
 By comparing our results with the experimental findings we give a
 prediction for the quantum numbers of these newly observed
 subatomic particles.

We use a well-defined procedure that was developed and utilized extensively by
HSC in extracting excited states of light mesons~\cite{Dudek:2009qf,Dudek:2010wm,Dudek:2010ew}, mesons containing charm
quarks~\cite{Dudek:2007wv,Liu:2012ze,Moir:2013ub}, light and strange baryons
~\cite{Edwards:2011jj, Edwards:2012fx}, as well as charm
baryons~\cite{Padmanath:2013zfa,Padmanath:2015jea,Padmanath:2013bla,Padmanath:2014bxa}. This method has the following important ingredients:

\noindent{\textbf{A. Anisotropic lattice:}}
We use a set of anisotropic gauge field configurations, where dynamics
of light and strange quarks are included.  The extended
time direction and fine temporal lattice spacing ($a_t$) are very
helpful to obtain better resolution of the correlation functions,
which is crucial for the reliable extraction of excited states. We use the tree-level Symanzik-improved gauge action
along with an anisotropic clover fermion action with tree-level
tadpole-improved and three-dimensional stout-link smeared gauge
fields. Following are lattice details : size = $16^3\times 128$; $a_t 
\sim 0.035$ fm with an anisotropy of 3.5; $m_{\pi} \sim 391$ MeV;
and the number of configurations is 96. Further details of the actions 
can be found
in Refs.~\cite{Edwards:2008ja, Lin:2008pr}.  The charm quark mass is
tuned by equating the lattice mass of the meson $\eta_c$(1S) with its
physical mass. Details of the charm quark action
are given in Ref.~\cite{Liu:2012ze}.

\noindent{\textbf{B. Large set of interpolating fields:}} 
Hadron spectroscopy on the lattice proceeds through an investigation of the two 
point correlation functions between the 
hadron interpolating fields (operators). 
Because of the octahedral symmetry, interpolating fields on the lattice are not the same as their continuum counterparts, and one needs to construct these interpolating fields according to the reduction of continuum fields into various lattice irreducible representations (irreps), namely, $G_1$, $G_2$, and $H$ for baryons~\cite{Johnson:1982yq}. Physical states with spins 1/2 and 
3/2 can then be obtained only from the $G_1$ and $H$ irreps respectively, while spin-5/2 states are accessible from both the $H$ and the $G_2$ irreps ~\cite{Johnson:1982yq}.
Following Refs.~\cite{Basak:2005ir,Edwards:2011jj} we construct a large set of
  baryonic operators which is essential for the 
reliable extraction of excited states from lattice calculations. 
 These operators transform as irreps of SU(3)$_F$ symmetry 
for flavor, SU(4) symmetry for Dirac spin of 
quarks and double cover octahedral group $O^D_h$ of the
lattice.
 The flavor content of $\Omega_c (ssc)$ baryons
is similar to that of $\Omega_{cc} (ccs)$ baryons with the role of
$c$ and $s$ quark exchanged. Hence, the operator details for $\Omega_c^0$
baryons used in this work follow from Section IIB and Tables II
and III of 
 Ref.~\cite{Padmanath:2015jea}, with the interchange of $c$ and $s$ quarks.

\noindent{\textbf{C. Distillation method:} 
We employed a novel technique called
``distillation''~\cite{Peardon:2009gh}, which is a quark source
smearing technique that enables one to compute large correlation
matrices ($C_{ij}$) between a large basis of operators
 including nonlocal ones,
similar to those used in this calculation. 
 Here we
implement the method using the lowest 64 eigenvectors of
the discretized gauge-covariant Laplacian. 
The correlation matrices are built from four different source time
slices.

\noindent{\textbf{D. Variational analysis and spin identification:}}
We utilize a robust analysis procedure, developed by HSC, which is
based on the variational study of correlation matrices, $C_{ij}$.  
In this method, one solves a generalized
eigenvalue problem (GEVP) \cite{Michael:1985ne,Luscher:1990ck} of the form 
\beq C_{ij}(t)v_{j}^{n} =
\lambda_{n}(t,t_0)C_{ij}(t_0)v_{j}^{n}, 
\eeq{GEVP} 
where
$\lambda_{n}(t,t_0)$ is the $n$ th eigenvalue which is related to the
energy of the $n$ th excited state $E_n$ by 
\beq
\lim_{t-t_0\rightarrow\infty} \lambda_n(t,t_0) = e^{-E_n (t-t_0)} .
\eeq{GEVP} 
We choose an appropriate reference time slice $t_0$ in solving
 GEVP, such that it minimizes a $\chi^2$-like quantity as defined
in Ref.~\cite{Dudek:2007wv}. 
To associate a spin to an extracted energy
level we calculate the overlap factors $Z^n_i$ of an operator $O_i$
defined as $Z^n_i \equiv \langle n|O_i^{\dagger}|0\rangle$
to a state $n$ with energy $E_n$. 
These overlap factors carry a memory of the 
corresponding continuum interpolating field from which $O_i$ was derived and 
these factors can be obtained from the $n$ th eigenvector $v^{n}$ of the GEVP. 
This procedure is being
widely used by HSC in all of its spectrum calculations.
\bef[tbh]
\centering
\includegraphics*[scale=0.37]{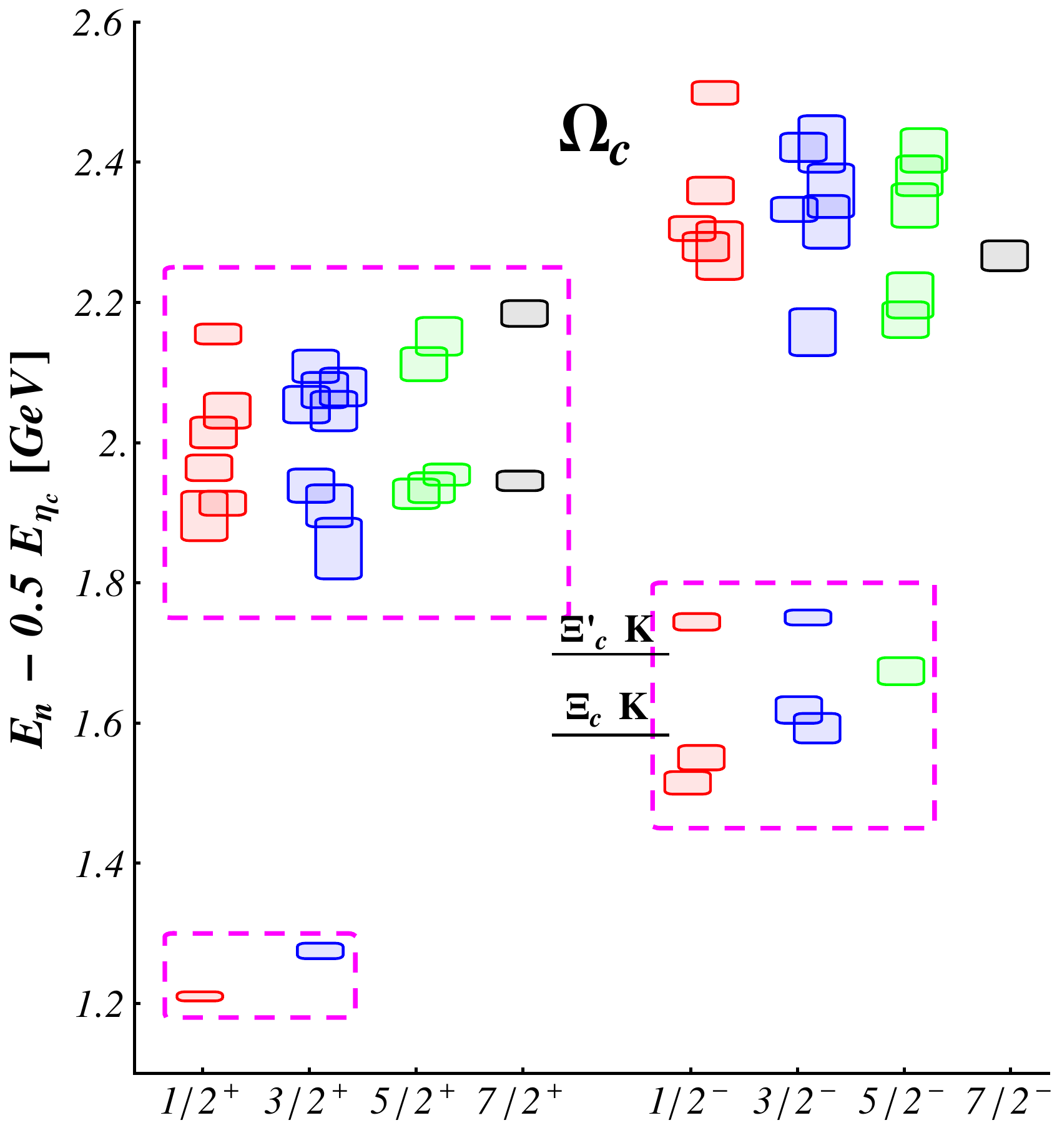}
\vspace*{-0.08in}
\caption{Spin identified spectra of $\Omega^{0}_{c}$ baryons. Here, spectrum is
  presented in terms of energy splittings of $\Omega^{0}_{c}$ baryons
  from $\eta_c$(1S) meson. Details of the plot
  are in the text.}
\eef{fig_Oc_spec_lattice}

\noindent{\bf{Results:}}
Following the above procedures, we are able to extract the energy spectrum of
$\Omega_{c}^{0}$ baryons with spin up to 7/2. In
~\fgn{fig_Oc_spec_lattice}, we show our results in terms of energy
splittings of $\Omega^{0}_{c}$ baryons from the mass of the $\eta_c(1S)$ meson. A
factor 1/2 is multiplied with $\eta_c$ mass to account for the
difference in the number of valence charm quarks in the baryon and meson.
In general, energy splittings with valence charm content subtracted
will have reduced uncertainties originating from the systematics of
the charm quark mass parameter in the lattice action and from the
ambiguity in the scale setting procedure. Positive and negative parity states are shown on the left- and right-hand sides of the figure, respectively.
The vertical height in each box represents $1\sigma$ uncertainty, which includes statistical
and systematic uncertainties from different fit ranges.
Throughout this Letter, we follow the color coding 
for extracted energy levels as follows : spin 1/2, red;
spin 3/2, blue; spin 5/2, green; and spin 7/2 black.  The two relevant scattering channels in this
calculation are $\Xi_cK$ and $\Xi^{\prime}_cK$ in $s$ wave. Their lattice
values are shown by horizontal black lines and are obtained from $\Xi_c$, $\Xi^{\prime}_c$ and $K$ masses calculated on these lattices. 
The states inside the magenta boxes are those with dominant
overlap to operators constructed purely out of the upper two
components of the quark spinor. Those are
referred to as the nonrelativistic operators. All other operators
are relativistic.  
It is interesting to see that the number of low lying excitations for
each spin agrees with the expectations based on the nonrelativistic
quark spins which implies a clear signature of SU(6)$\times$O(3)
symmetry in the spectra.  A similar SU(6)$\times$O(3) symmetric nature
of the spectra was also observed in light baryons~\cite{Edwards:2012fx} as
well as for doubly and triply heavy
baryons~\cite{Padmanath:2015jea,Padmanath:2013zfa}. It is to be noted
that in our variational analysis we have included both
nonrelativistic as well as relativistic operators, and still we
observe the above symmetry in the low lying spectra.

\bef[!t]
\centering
\includegraphics[scale=0.55]{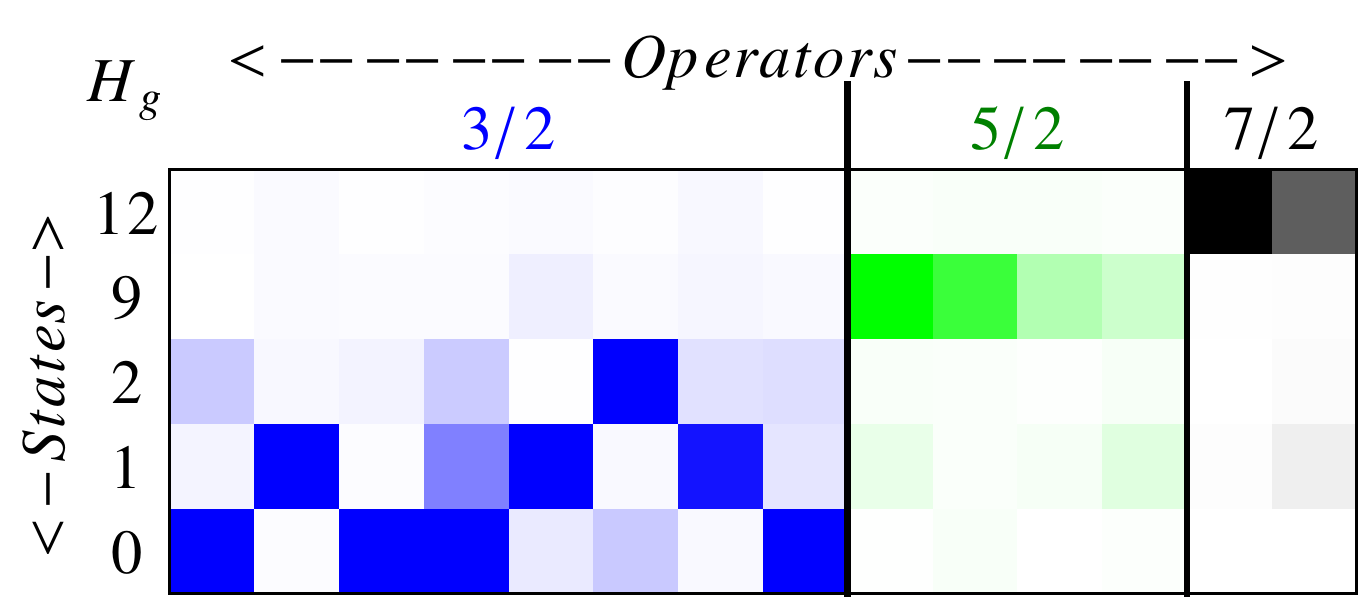}
\vspace*{-0.08in}
\caption{``Matrix" plot of $\tilde{Z}$ for a few selected operators onto a 
few spin identified lower energy levels in $H_g$ irrep.}
\eef{rel_Hg_t013}
\bef[tbh]
\centering
\vspace*{-0.1in}
\includegraphics[scale=0.55]{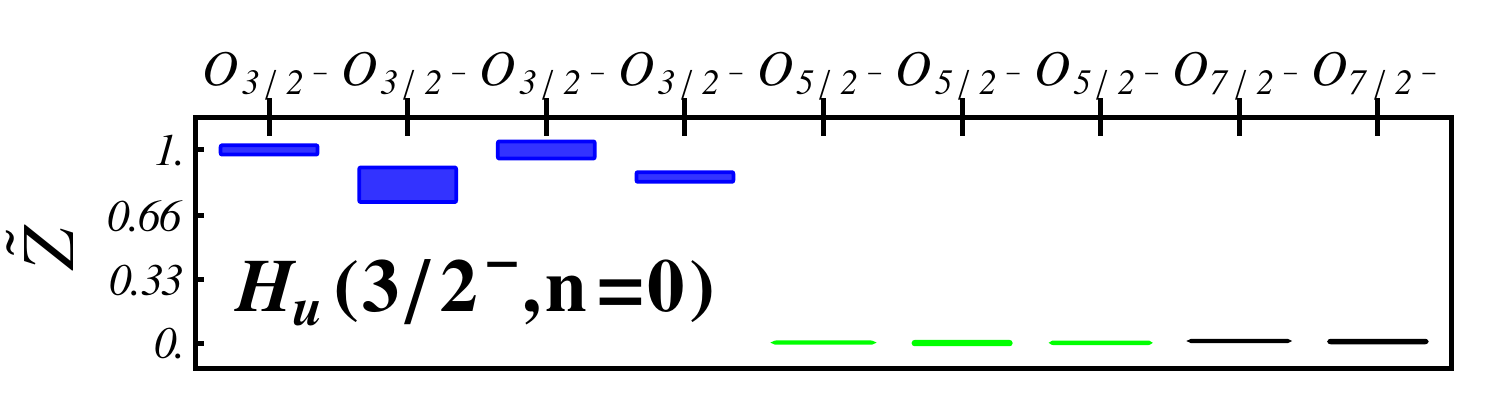}
\includegraphics[scale=0.555]{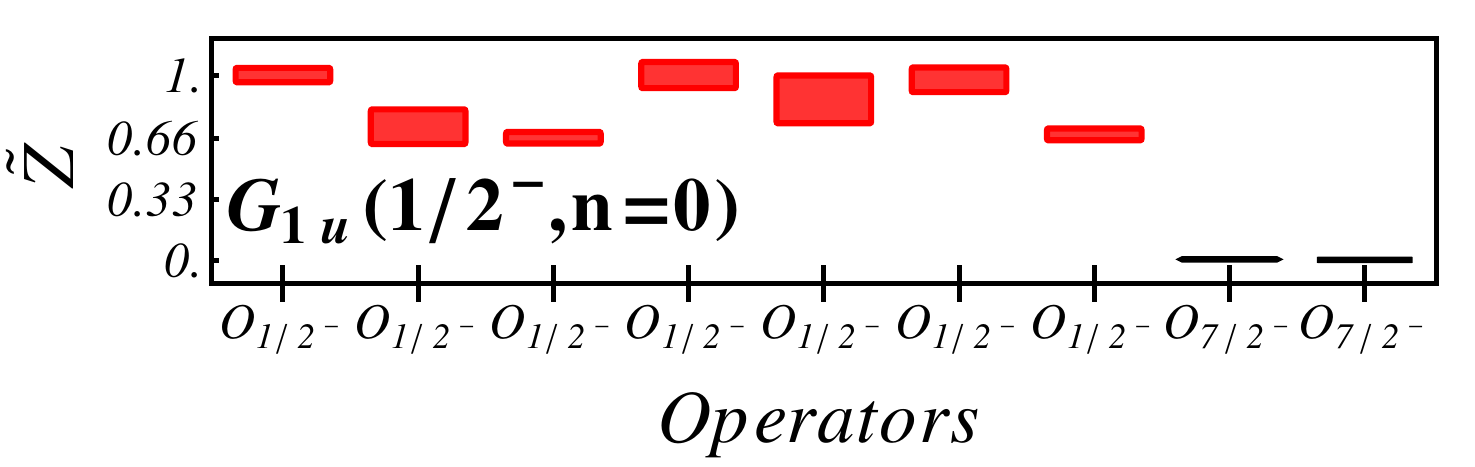}
\vspace*{-0.08in}
\caption{Histogram plot of $\tilde{Z}$ for the lowest levels in $H_u$ and $G_{1u}$ irreps for a few selected operators.}
\eef{spinhist}

Next, we briefly describe the procedure followed in assigning the spin
of an extracted energy level leading to the spin identified
spectra shown in ~\fgn{fig_Oc_spec_lattice}. To explain it, we choose
irreps $H_g$, $H_u$ and $G_{1u}$ (subscripts $g$ and 
$u$ refer to positive and negative parity, respectively), and show below how a particular
energy level that is associated with operators from any of these irreps
can be assigned a spin.  
In ~\fgn{rel_Hg_t013}, we show a
representative matrix plot of the normalized overlap factors,
$\tilde{Z} = \frac{Z^n_i}{max[Z_i^n]}$, for a few selected operators
on to a few of the lower energy levels in the $H_g$ irrep, where the continuum $3/2^+$, $5/2^+$, and $7/2^+$ states appear. We follow
the same color coding as above, while the darkness of the pixel
is linearly related to the magnitude of the normalized overlap.  From
this figure, one can clearly associate the states labeled as 0, 1, and 2 with spin 3/2, the state 9 with
spin 5/2, and the state 12 with spin 7/2.  In order to further demonstrate
the robustness of the procedure, in
\fgn{spinhist} we present a representative histogram plot of $\tilde{Z}$ values 
from two different irreps. In the top plot,
on the $x$ axis we show various operators in the $H_u$ irrep with
their continuum spins and the $y$ axis shows their $\tilde{Z}$ values to a particular energy excitation. This also shows that
this energy excitation represents a spin-$3/2^{-}$ state as it is
saturated predominantly from operators that have spin $3/2^{-}$ in the
continuum. 
In the bottom plot, we show a similar observation in the
$G_{1u}$ irrep for an energy excitation that we found to be a spin-$1/2^{-}$ state. For spin-5/2 and 7/2 states, their $\tilde{Z}$ values need 
to be compared among different irreps~\cite{Edwards:2011jj, Edwards:2012fx,Padmanath:2013zfa,Padmanath:2015jea}.
Spin identifications for all other energy excitations
are performed with the same rigor.

\bef
\centering
\includegraphics[scale=0.40]{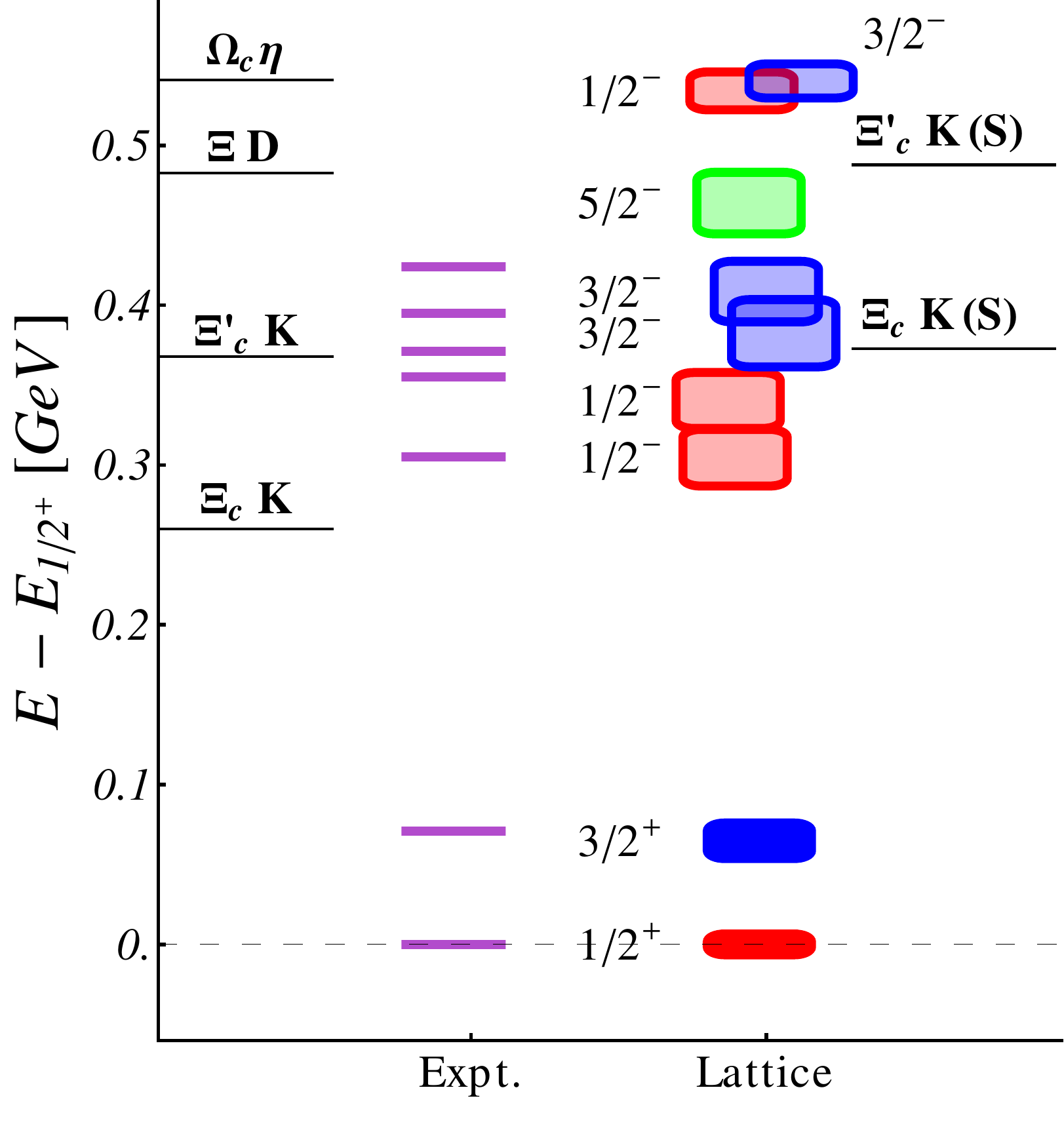}
\vspace*{-0.08in}
\caption{Comparison plot between experimental and lattice results of
  $\Omega_c$ baryons.}
\eef{LatVsExpt}
With confidence in our procedure for the extraction of energy levels
and their spin identification, we finally present our main result.  In
\fgn{LatVsExpt}, we show a comparison plot between the extracted
lattice energy levels with those from the recently observed $\Omega_c^{0}$
baryons~\cite{Aaij:2017nav} along with the
previously known two other $\Omega_c^{0}$ baryons~\cite{PDG}. 
The relevant continuum
scattering thresholds are presented on the left-hand side and the
noninteracting scattering energies as obtained on these lattices are
shown on the right-hand side.  It can be seen that our lattice estimate for the
hyperfine splitting between spin-3/2 and spin-1/2 baryons is well in
agreement with experiment. The most interesting observation
from this comparison is the fact that we observe exactly five energy
excitations in the same energy region above the $3/2^+$ state. Two other
states are above the scattering levels and thus need to be studied with more 
rigor.  It is also very satisfying to see that among the
five new excitations, four are matching with our lattice results. The
only remaining excitation, which we assign to be a spin-$5/2^-$
baryon, can possibly be identified to the remaining higher lying
experimental candidate. We would like to point out that these results
are prediction and not postdiction, as preliminary results of these were
already presented at the Charm-2013, 2015 and Lattice-2014 conferences~\cite{Padmanath:2013bla,Padmanath:2014bxa,Padmanath:2015bra}. It is to be noted that
most other nonlattice calculations~\cite{Ebert:2007nw,Ebert:2011kk,Garcilazo:2007eh,Valcarce:2008dr,Roberts:2007ni,Vijande:2012mk,Yoshida:2015tia,Shah:2016nxi,Bagan:1992tp,Huang:2000tn,Wang:2009cr,Chen:2015kpa,Chiladze:1997ev}
on $\Omega_c^0$ baryons predicted seven
levels in this region. In \tbn{summary_table}, we summarize the
comparison between experiments and this lattice calculation (called L1), where we denote the $i$ th energy level of $\Omega_c^{0}$
by $\Omega_c^{0,i}$, while $\Delta E$ is the energy difference from
 the ground state ($\Omega_c^{0,0}$).
\bet[h]
\centering
\begin{tabular}{c|| cc || ccc }
\hline
Energy & \multicolumn{2}{c||}{Experiment} & \multicolumn{3}{c}{Lattice} \\ \cline{2-6}
splittings ($\Delta E$) & $\Delta E$  & $J^{P}$   & \multicolumn{2}{c}{$\Delta E$ (MeV)} & $J^{P}$ \\
&(MeV)&\cite{PDG}&L1&L2&
\\\hline
$E_{\Omega_c^{0,0}} - {1\over 2} E_{\eta_c}$ &  1203(2)  & $1/2^+$ & 1209(7) & 1200(10) & $1/2^+$\\
$E_{\Omega_c^{0,1}} - E_{\Omega_c^{0,0}}$    & 70.7(1)   & $3/2^+$ & 65(11)  & 68(14)  & $3/2^+$ \\
$E_{\Omega_c^{0,2}} - E_{\Omega_c^{0,0}}$    & 305(1)    & ?       & 304(17) & 319(19) & $1/2^-$\\
$E_{\Omega_c^{0,3}} - E_{\Omega_c^{0,0}}$    & 355(1)    & ?       & 341(18) &         & $1/2^-$ \\
$E_{\Omega_c^{0,4}} - E_{\Omega_c^{0,0}}$    & 371(1)    & ?       & 383(21) &  & $3/2^-$\\
$E_{\Omega_c^{0,5}} - E_{\Omega_c^{0,0}}$    & 395(1)    & ?       & 409(19) & 403(21)      & $3/2^-$\\
$E_{\Omega_c^{0,6}} - E_{\Omega_c^{0,0}}$    & 422(1)    & ?       & 464(20) &         & $5/2^-$\\\hline
\hline
\end{tabular}
\caption{Comparison of energy splittings of $\Omega_c^{0}$ baryons
  between experimental and lattice results. $\Omega_c^{0,i}$ represents
the $i$ th energy level.}
\eet{summary_table}
 From the results shown in \fgn{LatVsExpt} and \tbn{summary_table}, we conclude  that the spin-parity quantum
 numbers of these newly discovered particles are as follows: $\Omega_c(3000)^0$
 and $\Omega_c(3050)^0$ have spin-parity $J^P = 1/2^{-}$, the states
 $\Omega_c(3066)^0$ and $\Omega_c(3090)^0$ have $J^P = 3/2^{-}$, while
 $\Omega_c(3119)^0$ is possibly a $5/2^{-}$ state. A similar assignment of spin and parity has also recently been made in a potential model calculation~\cite{Karliner:2017kfm}.

To strengthen our findings, we perform another lattice calculation (called L2) with a
totally different set of lattice parameters. 
We use three
dynamical 2+1+1 flavors HISQ lattice ensembles generated by the MILC
Collaboration~\cite{Bazavov:2012xda} : $24^3 \times 64$, $32^3 \times
96$, and $48^3 \times 96$ lattices with lattice spacings $~\sim$ 0.12,
0.09, and 0.06 fm, respectively. For the valence quark propagators,
we use overlap action~\cite{Neuberger:1997fp}. The details of this
lattice set up, charm and strange mass tuning are given in
Refs.~\cite{Basak:2012py,Basak:2013oya}. On these ensembles,  we calculate two point correlation functions of
$\Omega_c^0$ baryons using conventional local spin-1/2 and 3/2 operators and
extract the respective lowest states for both the parities. In
\fgn{second_lat_result}, we show results from this calculation again
as mass splittings from the ground state ($1/2^{+}$). We also perform continuum extrapolations using
 $\mathcal{O}(a^2)$ and $\mathcal{O}(a^3)$ forms in lattice spacing, $a$. The 
$1\sigma$ error bars from $\mathcal{O}(a^3)$ fittings 
are shown by the shaded regions. In Table I, in the column L2 we show these results which include both statistical as well as all systematic errors. 
It is quite encouraging to
see that lattice results from two completely different setups are
consistent with each other, and this ensures again the
robustness of the spin assignment procedure utilized
in the first calculation. It is to be noted that the results obtained in the L2 calculation rely on conventional single exponential fits of the two point correlation function. Hence, while results for spin-$1/2^+$ and -$3/2^+$ states are reliable, it is difficult to extract energy levels of states reliably whose energies are close by. In that case, one obtains a single energy level as a mixture of the two. This is indeed what we observed in the L2 results for $1/2^-$ states, which is in the middle of two states obtained in the L1 calculation.

\bef[t!]
\centering
\vspace*{-0.2in}
\includegraphics[scale=0.42]{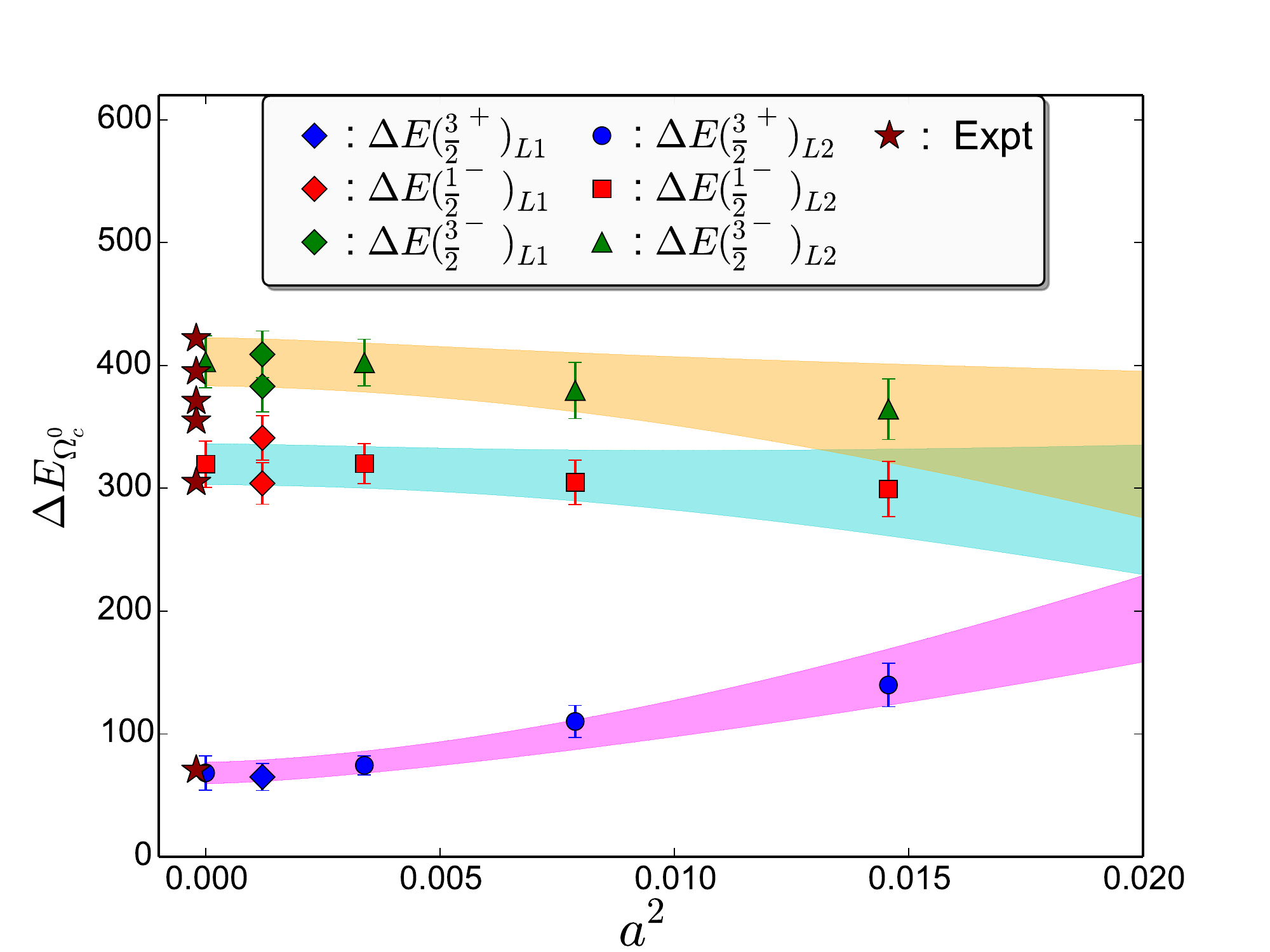}
\vspace{-0.08in}
\caption{Mass splittings of the lattice energy levels from the ground
  state of $1/2^{+}$ $\Omega_c^0$ baryon, where $\Delta E(J^P) \equiv E(J^P) - E(1/2^{+})$. L1 and L2 represent first and second lattice calculations.}
\eef{second_lat_result}

We now discuss the scattering channels relevant to our
calculation.  For the lowest three states the only
possible strong decay channel is $\Xi^+K^-$, whereas the higher two
levels can decay into $\Xi^+K^-$ and $\Xi^{\prime +}K^-$.  Owing to the
heavy pion mass ($m_{\pi}\sim391$ MeV),  
scattering levels $\Xi_c^+K^-$ and $\Xi_c^{\prime +}K^-$
in the $s$ wave, as measured on our lattice, appear at 373 and 488 MeV, 
respectively, above the ground state, as shown in \fgn{LatVsExpt}. Both
of the extracted spin-$1/2^{-}$ states are below these energy
thresholds. On the other hand, spin-$3/2^{-}$ and spin-$5/2^{-}$ states
can decay into $\Xi_c^+K^-$ only via $d$ wave.  However, the
corresponding noninteracting lattice scattering energies lie
significantly above these excitations.  
Considering the narrow width of the observed resonances \cite{Aaij:2017nav}
and lattice positions of the scattering channels, as discussed above,  
we  believe that the
single hadron approximation in our calculation, where we have
neglected multiparticle operators, will have negligible effects on
the energy excitations that we have extracted.

We would also like to point out possible uncertainties in this
calculation. The main uncertainties are from the discretization of the heavy
charm quark mass. As mentioned previously, we believe that this
uncertainty gets reduced by taking appropriate energy differences
 where the effects of the valence charm quark
content is subtracted out.  The agreement between lattice and
experimental values in the hyperfine splitting between spin-3/2 and
spin-1/2 baryons, which is known for its strong discretization
artifacts, also justifies the above claim. 
Furthermore, consistency between our two lattice investigations (L1 and L2) with entirely different systematics confirms that the discretization effects on the mass splittings in the first calculation are indeed under control. 
The effects from the unphysical light quark mass and small lattice volume 
are expected to be smaller in $\Omega_c$ baryons than those for light baryons, as the former do not have light valence $u$ and $d$ quarks. Our lattice value of the spin-1/2 ground
state matches with its experimental value, which further provides
confidence to this view. The relative spin ordering of these energy
excitations that we assigned here is expected to be unaffected by 
future lattice calculations with more realistic physical parameters.

\noindent{\bf{Conclusions: }}
In this Letter, we present detailed results from the first
nonperturbative calculation on the excited state spectroscopy of
$\Omega^{0}_{c}$ baryons with spin up to 7/2 and for positive as well
as negative parity.  Results from this work have direct relevance to
the five $\Omega^{0}_{c}$ resonances recently discovered by the LHCb
Collaboration. We predict the quantum number of these energy
excitations as the following: $\Omega_c(3000)^0$ and $\Omega_c(3050)^0$
have spin-parity $J^P = 1/2^{-}$, the states $\Omega_c(3066)^0$ and
$\Omega_c(3090)^0$ have $J^P = 3/2^{-}$, whereas $\Omega_c(3119)^0$
possibly is a $5/2^{-}$ state. An elaborate and well-established
lattice method is followed for extracting these energy levels and in
identifying their spins. We cross-check these results by performing
another lattice calculation with a completely different setup and with
 better control over systematics. The spin-parity quantum number assigned
to these newly observed states is expected to be unaffected by any
future lattice calculation with much improved control over the
systematic uncertainties.

{\bf{Acknowledgements:}} We thank our colleagues within the Hadron
Spectrum Collaboration.  We are thankful to the MILC Collaboration 
and, in particular, to S. Gottlieb for providing us with the HISQ lattices.
Computations are carried out on the Blue Gene/P of the ILGTI in TIFR,
and on the Gaggle cluster of the Department of Theoretical Physics,
TIFR. Chroma~\cite{Edwards:2004sx} and 
QUDA~\cite{Clark:2009wm,Babich:2010mu} software are used for this
calculation. N. M. would like to thank A. Dighe and P. Junnarkar,  
and M. P. would like
to thank S. Collins for discussions. M. P. also acknowledges support
from the Austrian Science Fund FWF:I1313-N27 and the Deutsche
Forschungsgemeinschaft under Grant No.SFB/TRR 55.


\begin{thebibliography}{99}
\bibitem{PDG} C. Patrignani {\it et al.}, Chin. Phys. C 40, 100001 (2016). 
\bibitem{Aaij:2017nav} 
  R.~Aaij {\it et al.} (LHCb Collaboration),
  Phys.\ Rev.\ Lett. \textbf{118}, 182001 (2017).



\bibitem{Ebert:2007nw} D.~Ebert, R.~N.~Faustov, and V.~O.~Galkin,
Phys.\ Lett.\ B \textbf{659}, 612 (2008).
\bibitem{Ebert:2011kk} D.~Ebert, R.~N.~Faustov, and V.~O.~Galkin,
Phys.\ Rev.\ D \textbf{84}, 014025 (2011).


\bibitem{Garcilazo:2007eh} H.~Garcilazo, J.~Vijande, and A.~Valcarce,
J.\ Phys.\ G \textbf{34}, 961 (2007).
\bibitem{Valcarce:2008dr} A.~Valcarce, H.~Garcilazo, and J.~Vijande,
Eur.\ Phys.\ J.\ A \textbf{37}, 217 (2008).


\bibitem{Roberts:2007ni} W.~Roberts and M.~Pervin,
Int.\ J.\ Mod.\ Phys.\ A \textbf{23}, 2817 (2008).


\bibitem{Vijande:2012mk} J.~Vijande, A.~Valcarce, T.~F.~Carames, and H.~Garcilazo, 
Int.\ J.\ Mod.\ Phys.\ E \textbf{22}, 1330011 (2013).


\bibitem{Yoshida:2015tia} T.~Yoshida E.~Hiyama, A.~Hosaka, M.~Oka, and K.~Sadato, 
Phys.\ Rev.\ D \textbf{92}, 114029 (2015).

\bibitem{Shah:2016nxi} 
  Z.~Shah, K.~Thakkar, A.~K.~Rai, and P.~C.~Vinodkumar,
  Chin.\ Phys.\ C {\bf 40}, 123102 (2016).


\bibitem{Bagan:1992tp} E.~Bagan, M.~Chabab, H.~G.~Dosch, and S.~Narison,
Phys.\ Lett.\ B \textbf{287}, 176 (1992).

\bibitem{Huang:2000tn} C.~S.~Huang, A.~l.~Zhang, and S.~L.~Zhu,
Phys.\ Lett.\ B \textbf{492}, 288 (2000).


\bibitem{Wang:2009cr} Z.~G.~Wang,
Phys.\ Lett.\ B \textbf{685}, 59 (2010).


\bibitem{Chen:2015kpa} H.~X.~Chen {\it et al.}, 
Phys.\ Rev.\ D \textbf{91}, 054034 (2015).



\bibitem{Chiladze:1997ev} G.~Chiladze and A.~F.~Falk,
Phys.\ Rev.\ D \textbf{56}, R6738 (1997).


\bibitem{Dudek:2007wv} 
  J.~J.~Dudek, R.~G.~Edwards, N.~Mathur, and D.~G.~Richards,
  Phys.\ Rev.\ D {\bf 77}, 034501 (2008).

\bibitem{Liu:2012ze} 
  L.~Liu {\it et al.}  [Hadron Spectrum Collaboration],
  JHEP {\bf 1207}, 126 (2012).

\bibitem{Moir:2013ub} 
  G.~Moir, M.~Peardon, S.~M.~Ryan, C.~E.~Thomas, and L.~Liu,
  JHEP {\bf 1305}, 021 (2013).


\bibitem{Mathur:2002ce} 
  N.~Mathur, R.~Lewis, and R.~M.~Woloshyn,
  Phys.\ Rev.\ D {\bf 66}, 014502 (2002).

\bibitem{Lewis:2001iz} 
  R.~Lewis, N.~Mathur, and R.~M.~Woloshyn,
  Phys.\ Rev.\ D {\bf 64}, 094509 (2001).

\bibitem{Durr:2012dw} 
  S.~Durr,  G.~Koutsou, and T.~Lippert,
  Phys.\ Rev.\ D {\bf 86}, 114514 (2012).

\bibitem{Basak:2012py} 
  S.~Basak, S.~Datta, M.~Padmanath, P.~Majumdar, and N.~Mathur,
  PoS LATTICE {\bf 2012}, 141 (2012), arXiv:1211.6277.

\bibitem{Basak:2013oya} 
  S.~Basak, S.~Datta, A.~T.~Lytle, M.~Padmanath, P.~Majumdar, and N.~Mathur,
  PoS LATTICE {\bf 2013}, 243 (2014), arXiv:1312.3050.

\bibitem{Namekawa:2013vu} 
  Y.~Namekawa {\it et al.} [PACS-CS Collaboration],
  Phys.\ Rev.\ D {\bf 87}, 094512 (2013).

\bibitem{Brown:2014ena} 
  Z.~S.~Brown,  W.~Detmold, S.~Meinel, and K.~Orginos,
  Phys.\ Rev.\ D {\bf 90}, 094507 (2014).

\bibitem{Bali:2015lka} 
  P.~Perez-Rubio, S.~Collins, and G.~S.~Bali,
  Phys.\ Rev.\ D {\bf 92}, 034504 (2015).

\bibitem{Padmanath:2013zfa} 
  M.~Padmanath, R.~G.~Edwards, N.~Mathur, and M.~Peardon,
  Phys.\ Rev.\ D {\bf 90}, 074504 (2014).

\bibitem{Padmanath:2015jea} 
  M.~Padmanath, R.~G.~Edwards, N.~Mathur, and M.~Peardon,
  Phys.\ Rev.\ D {\bf 91}, 094502 (2015).


\bibitem{Padmanath:2013bla} 
  M.~Padmanath, R.~G.~Edwards, N.~Mathur, and M.~Peardon,
   arXiv:1311.4806.

\bibitem{Padmanath:2014bxa} 
  M.~Padmanath, R.~G.~Edwards, N.~Mathur, and M.~J.~Peardon,
  PoS LATTICE {\bf 2014}, 084 (2015), arXiv:1410.8791.

\bibitem{Padmanath:2015bra} 
  M.~Padmanath and N.~Mathur,
  arXiv:1508.07168.



\bibitem{Dudek:2010wm} 
  J.~J.~Dudek, R.~G.~Edwards, M.~J.~Peardon, D.~G.~Richards, and C.~E.~Thomas,
  Phys.\ Rev.\ D {\bf 82}, 034508 (2010).
%
\bibitem{Dudek:2009qf} 
  J.~J.~Dudek, R.~G.~Edwards, M.~J.~Peardon, D.~G.~Richards, and C.~E.~Thomas,
  Phys.\ Rev.\ Lett.\  {\bf 103}, 262001 (2009).
%
\bibitem{Dudek:2010ew} 
  J.~J.~Dudek, R.~G.~Edwards, M.~J.~Peardon, D.~G.~Richards, and C.~E.~Thomas,
  Phys.\ Rev.\ D {\bf 83}, 071504 (2011).


\bibitem{Edwards:2011jj} 
  R.~G.~Edwards, J.~J.~Dudek, D.~G.~Richards, and S.~J.~Wallace,
  Phys.\ Rev.\ D {\bf 84}, 074508 (2011).


\bibitem{Edwards:2012fx} 
  R.~G.~Edwards,  N.~Mathur, D.~G.~Richards, and S.~J.~Wallace,
  [Hadron Spectrum Collaboration],
  Phys.\ Rev.\ D {\bf 87}, 054506 (2013).


\bibitem{Edwards:2008ja} 
  R.~G.~Edwards, B.~Joo, and H.~-W.~Lin,
  Phys.\ Rev.\ D {\bf 78}, 054501 (2008).

\bibitem{Lin:2008pr} 
  H.~-W.~Lin {\it et al.} (Hadron Spectrum Collaboration),
  Phys.\ Rev.\ D {\bf 79}, 034502 (2009).

\bibitem{Johnson:1982yq} 
  R.~C.~Johnson,
  Phys.\ Lett.\  {\bf B 114}, 147 (1982).

\bibitem{Basak:2005ir}
  S.~Basak {\it et al.} (Lattice Hadron Physics (LHPC) Collaboration),
  Phys.\ Rev.\ D {\bf 72}, 074501 (2005).


\bibitem{Peardon:2009gh} 
  M.~Peardon {\it et al.}, (Hadron Spectrum Collaboration),
  Phys.\ Rev.\ D {\bf 80}, 054506 (2009).


\bibitem{Michael:1985ne} 
  C.~Michael,
  Nucl.\ Phys.\ {\bf B259}, 58 (1985).

\bibitem{Luscher:1990ck} 
  M.~Luscher and U.~Wolff,
  Nucl.\ Phys.\ {\bf B339}, 222 (1990).



\bibitem{Karliner:2017kfm} 
  M.~Karliner and J.~L.~Rosner,
  Phys.\ Rev.\ D {\bf 95}, 114012 (2017).

\bibitem{Bazavov:2012xda} 
  A.~Bazavov {\it et al.} (MILC Collaboration),
  Phys.\ Rev.\ D {\bf 87}, 054505 (2013).

\bibitem{Neuberger:1997fp} 
  H.~Neuberger,
  Phys.\ Lett.\ B {\bf 417}, 141 (1998).

\bibitem{Edwards:2004sx} 
  R.~G.~Edwards {\it et al.} (SciDAC and LHPC and UKQCD Collaborations),
  Nucl.\ Phys.\ Proc.\ Suppl.\  {\bf 140}, 832 (2005).

\bibitem{Clark:2009wm} 
  M.~A.~Clark, R.~Babich, K.~Barros, R.~C.~Brower, and C.~Rebbi,
  Comput.\ Phys.\ Commun.\  {\bf 181}, 1517 (2010).


\bibitem{Babich:2010mu} 
  R.~Babich, M.~A.~Clark, and B.~Joo,
  arXiv:1011.0024.


\end{thebibliography}
\end{document}